\documentclass{article}
\usepackage{amsfonts,amssymb, amsmath}
\usepackage{relsize}
\usepackage[english]{babel}

\def\bq{ \begin{equation} }
\def\eq{ \end{equation} }
\def\ben{ \begin{eqnarray} }
\def\en{ \end{eqnarray} }

\begin{document}


\title{On bi-Hamiltonian formulation of the perturbed Kepler problem}

\author{ Yu. A. Grigoryev, A. V. Tsiganov\\
\it\small
St.Petersburg State University, St.Petersburg, Russia\\
\it\small e--mail: andrey.tsiganov@gmail.com, yury.grigoryev@gmail.com}

\date{}
\maketitle

\begin{abstract}
The perturbed Kepler problem is shown to be a  bi-Hamiltonian system  in spite of the fact that the graph of the Hamilton function is not a hypersurface of translation, which is against a necessary condition for the existence of the bi-Hamiltonian structure according to the Fernandes theorem. In fact,  both the initial and perturbed Kepler systems  are  isochronous systems and, therefore, the Fernandes theorem cannot be applied to them.
 \end{abstract}

\section{Introduction}
\setcounter{equation}{0}
Over the last  few years Magri's approach \cite{mag78} to integrability through bi-Hamiltonian structures had become one of the powerful methods of integrability of evolution equations applicable  in studying both finite and infinite dimensional dynamical systems.

The global, topological obstructions to the existence of a bi-Hamiltonian structure for a general completely integrable Hamiltonian system are discussed in  \cite{bog96,fern94,olver93,sm97}. Some counterexamples were given to show that  an existence of a bi-Hamiltonian structure is not always satisfied around a Liouville torus for a given Arnold-Liouville system. For instance, Fernandes \cite{fern94} and Olver \cite{olver93}  announced that the perturbed Kepler problem is a completely integrable system without a  bi-Hamiltonian formulation with respect to non-degenerate compatible Poisson structures in contrast with the initial Kepler problem.

 Below we explicitly present a few non-degenerate bi-Hamiltonian formulations of the perturbed Kepler problem using  the Bogoyavlenskij construction of a continuum of compatible Poisson structures for the isochronous Hamiltonian systems \cite{bog96}.

A bi-Hamiltonian vector field  is one which allows two Hamiltonian formulations
\bq\label{x-bi}
X=PdH=P'dK\,.
\eq
Here $P$ and $P'$ are the two compatible Poisson bivectors with vanishing Schouten brackets
\bq\label{meq}
[P,P]=[P,P']=[P',P']=0\,.
\eq
In generic cases bivectors $P$ and $P'$ could be degenerate and Hamiltonians  $H$ and $K$ could be functionally dependent. However, under an additional assumption one can construct a complete sequence of functionally independent first integrals   of $X$  \cite{bog96,mag78,mag97}.

The aim of this note is to present a bi-Hamiltonian formulation of the perturbed Kepler vector field $X$ defined by the Hamilton function
\bq\label{kepl-ham}
H=\dfrac{p_x^2+p_y^2+p_z^2}{2}-\dfrac{1}{r}+\dfrac{\epsilon}{2r^2}\,,\qquad r=\sqrt{x^2+y^2+z^2}\,,
\eq
and canonical  Poisson bivector
\bq\label{can-poi}
P=\left(
    \begin{array}{cc}
      0 & \mathrm I \\
      -\mathrm I & 0 \\
    \end{array}
  \right)\,.
\eq
The $r^{-2}$ correction can be seen as due to an asymmetric mass distribution of the attracting body (e.g., the gravitational attraction of the earth on a nearby orbiting artificial satellite) or
to non-Newtonian perturbation from the theory of general relativity (e.g., the motion of a particle in the Schwarzschild spherically symmetric solution of the Einstein equations), see \cite{vin}.

\section{Bi-Hamiltonian formulation of the perturbed Kepler problem}

Following to  \cite{bog96,fern94} we start with a discussion of the bi-Hamiltonian vector fields in terms of the action-angle variables.

According to the classical Arnold-Liouville theory  in the action-angle variables  $J_k$ and $ \omega_k$ a given  Hamiltonian vector field $X=PdH$ has the simple form
\bq\label{x-action}
X:\qquad \dot{J_k}=0\,,\qquad \dot{\omega}_k=\dfrac{\partial H}{\partial J_k}\,,\qquad k=1,\ldots,n.
\eq
Here $H=H(J_1,\ldots,J_n)$ is  a Hamilton function and the  Poisson bivector $P$ is  the canonical one
\bq\label{can-poi2}
P=\sum_{k=1}^n \dfrac{\partial}{\partial J_k}\wedge\dfrac{\partial}{\partial \omega_k}\,.
\eq
 The problem of  the existence of the action-angle variables  in the neighborhood of an orbit, of a level set or globally is discussed in \cite{du80}, see also the recent review \cite{mir14} and references within. We will look for a bi-Hamiltonian formulation only in the domain of definition of the action-angle variables.

 The vector field $X$  is called non-degenerate or anisochronous if the Kolmogorov condition for the Hessian matrix
\bq\label{non-deg}
\det\left| \dfrac{\partial^2 H(J_1,\ldots,J_n)}{\partial J_i\partial J_k}\right|\neq 0
\eq
is met almost everywhere in the given  action-angle coordinates. This condition implies that the dense subsets of the invariant $n$-dimensional tori of $X$ are closures of trajectories.

In \cite{bog96} Bogoyavlenskij proposed a  complete classification of the  invariant Poisson structures
for  non-degenerate and  degenerate Hamiltonian systems, see Theorem 1 and Theorem 8, respectively.

Let us consider one trivial example of the generic Bogoyavlenskij construction for the  degenerate or isochronous  Hamiltonian system. If in the domain of definition of the action-angle variables we have some nonzero derivative
\[
a=\dfrac{\partial H}{\partial J_m}\neq 0\,,
\]
we can  make the following canonical transformation
\bq\label{trans1}
\begin{array}{l}
\tilde{J}_k=J_k\,,\qquad  \tilde{\omega}_k=\omega_k -\dfrac{\partial H}{\partial J_k}\, a^{-1}\,\omega_m\,,\qquad  k\neq m
\\
\\
\tilde{J}_m=H\,,\qquad  \tilde{\omega}_m=a^{-1}\omega_m\,.
\end{array}
 \eq
This canonical transformation  does not add new singularities to the  initial action-angle variables and reduces the Hamiltonian to the simplest form
\[H=\tilde{J}_m.\]
It allows us to construct   bi-Hamiltonian formulation of the  initial vector field $X$ with two functionally dependent Hamiltonians
\bq\label{part1-sol}
H=\tilde{J}_m\qquad\mbox{and}\qquad K=g(\tilde{J}_m)\,,
\eq
but with the non-degenerate second Poisson bivector
\bq\label{part1-poi}
P'=
\sum_{k\neq m}^n\beta_k(\tilde{J}_k)\,\dfrac{\partial}{\partial \tilde{J}_k}\wedge\dfrac{\partial}{\partial \omega_k}+
 \left(\dfrac{dg}{d\tilde{J}_m}\right)^{-1}\,\dfrac{\partial}{\partial\tilde{ J}_m}\wedge\dfrac{\partial}{\partial \tilde{\omega}_m}
 \,,
\eq
where $\beta_{k}(\tilde{J}_k)$ are arbitrary nonzero functions and $g(\tilde{J_m})$ is such that $g'\neq 0$.  In this case the eigenvalues of the corresponding recursion operator $N=P'P^{-1}$ are  integrals of motion only, see examples of this type bi-Hamiltonian formulations of the Kepler problem in \cite{bog96,marm90,sm97}.

In fact the Bogoyavlenskij theorems are much more powerful, however, in the following discussion, we need only this particular case.

\subsection{Action-angle variables for the perturbed Kepler problem}
Let us introduce the  spherical coordinates $r, \theta$ and $\phi$
\[
r=\sqrt{x^2+y^2+z^2}\,,\qquad \theta=\arctan\left(\dfrac{y}{x}\right)\,,\qquad \phi=\arccos\left(\dfrac{z}{\sqrt{x^2+y^2+z^2}}\right)
\]
which are  the radius, longitude and azimuth, respectively.
In order to describe the corresponding momenta we can use the so-called Mathieu generating function
\[F = p_x r \sin\phi\cos\theta + p_y r\sin\phi\sin\theta + p_z r\cos\phi\,,\]
so that
\[
p_r=\dfrac{\partial F}{\partial r}\,, \qquad p_\theta=\dfrac{\partial F}{\partial \theta}\,,\qquad
p_\phi=\dfrac{\partial F}{\partial \phi}\,.
\]
In the spherical variables Hamiltonian $H$ (\ref{kepl-ham}) takes the form
\[
H=\dfrac{1}{2}\left(p_r^2+\dfrac{p_{\theta}^2}{r^2}+\dfrac{p_\phi^2}{r^2\sin^2\theta}\right)-\dfrac{1}{r}+\dfrac{\epsilon}{2r^2}\,,
\]
such that the Hamilton-Jacobi equation $H=h$ has an additive separable solution
\[S=S_r(r)+S_\theta(\theta)+S_\phi(\phi)\]
that allows us to  introduce  the action-angle variables. Let us begin with the definition of two additional commuting  integrals  of motion
\[
\ell^2=p^2_\theta+\dfrac{p_\phi^2}{\sin^2\theta}\,,\qquad
m=p_\phi
\]
which are  the total angular momentum and the component of the angular momentum along the polar axis. Then  for $H=h<0$ we can calculate the action variables
\bq\label{a-var1}
\begin{array}{l}
J_\phi=\dfrac{1}{2\pi}\,\mathlarger{\oint} p_\phi d\phi=m\,,\\ \\
J_\theta=\dfrac{1}{2\pi}\,\mathlarger{\oint} p_\theta d\theta
=\dfrac{1}{2\pi}\,\mathlarger{\oint}\sqrt{\ell^2-\dfrac{m^2}{sin^2\theta}\,}\,d\theta
=\ell-m\,,\\ \\
J_r=\dfrac{1}{2\pi}\,\mathlarger{\oint} p_r dr
=\dfrac{1}{2\pi}\,\mathlarger{\oint}\sqrt{2h+\dfrac{2}{r}-\dfrac{\ell^2+\epsilon}{r^2}\,} \,dr
=\dfrac{1}{\sqrt{-2h}}-\sqrt{\ell^2+\epsilon}
\end{array}
\eq
using the standard integration method  \cite{born}. Then the  corresponding angle variables can be obtained from the Jacobi equations
\bq\label{a-var2}
\begin{array}{l}
\omega_r=\dfrac{\partial S}{\partial J_r}=-rp_r\sqrt{-2h}+\arccos\dfrac{1+2rh}{\sqrt{1+2h(\ell^2+\epsilon)}}\,,\\
\\
\omega_\theta=\dfrac{\partial S}{\partial J_\theta}=\dfrac{\ell}{\sqrt{\ell^2+\epsilon}}\left(\arccos\dfrac{1-\frac{\ell^2+\epsilon}{r}}{\sqrt{1+2h(\ell^2+\epsilon)}}-\omega_r\right)
+\arcsin\dfrac{\ell\cos\theta}{\sqrt{\ell^2-m^2}}\,,\\
\\
\omega_\phi=\dfrac{\partial S}{\partial J_\phi}=\omega_\theta+\phi+\arcsin\dfrac{m\cot\theta}{\sqrt{\ell^2-m^2}}\,.
\end{array}
\eq
The Hamiltonian $H$ (\ref{kepl-ham}) in these action-angle  variables takes the form
\bq\label{kepl-ham1}
H=-\dfrac{1}{2\left(J_r+\sqrt{(J_\theta+J_\phi)^2+\epsilon}\right)^2}\,.
\eq
It is easy to prove that the graph of $H$ (\ref{kepl-ham1})  is not a hypersurface of translation in the  action variables (\ref{a-var1}) \cite{fern94} in contrast with the initial Kepler problem at $\epsilon=0$.

The perturbed Kepler problem at $\epsilon\neq 0$ and the unperturbed Kepler problem  at  $\epsilon= 0$ are degenerate or isochronous systems
\[
\det\left| \dfrac{\partial^2 H(J_1,\ldots,J_n)}{\partial J_i\partial J_k}\right|= 0
\]
with well-defined derivative for $H=h<0$
\[
a=\dfrac{\partial H}{\partial J_r}=-(-2h)^{3/2}\,.
\]
According to the Bogoyavlenskij theorem it allows us to get bi-Hamiltonian formulations of these systems in the domain of definition of the action-angle variables (\ref{a-var1},\ref{a-var2}).

\subsection{Delaunay type variables}
The action coordinates play an important role in classical dynamics because of their adiabatic invariance, i.e.\ invariance under infinitesimally slow perturbations.  In the Kepler problem, there are few well-known types of orbits and, therefore, there are few types of the action-angle variables associated with different orbits. For instance, the Delaunay elements are valid only in the domain in phase space where there are  the elliptic orbits \cite{mar03}. On the other hand the two families  of the Poincar\'{e} variables, which are the action-angle  coordinates in the phase space of the Kepler problem, in the neighborhood of horizontal circular motions when eccentricities and inclinations are small \cite{fej13}. There are also Delaunay-similar elements,  Poincar\'{e}-similar elements and some other action-angle variables, which are well-defined in the neighborhood of different orbits.

For the perturbed Kepler problem we can also introduce the Delaunay type variables
\bq\label{del-var}
\begin{array}{lll}
J_1=J_\phi\,,\quad & J_2= J_\phi+J_\theta\,,\quad& J_3=J_r+\sqrt{\ell^2+\epsilon\,}\,,\\
\\
\omega_1= \omega_\phi-\omega_\theta\,,\quad & \omega_2=\omega_\theta-\dfrac{\ell}{\sqrt{\ell^2+\epsilon\,}}\,\omega_r\,,\quad&
\omega_3=\omega_r\,.
\end{array}
\eq
 Recall that the Delauney variables have a geometrical meaning directly related to the description of the orbits and their variations are much more significant for the astronomers than those of Cartesian or spherical variables \cite{fej13,mar03, vin}. For $\epsilon=0$  variables $J_k,\omega_k$  (\ref{del-var}) coincide with the  classical Delaunay elements $ (\mathrm{l, g, h,L,G, H})$:
\[\begin{array}{ll}
J_3\equiv \mathrm L=\sqrt{a}\,,\qquad& \omega_3\equiv \mathrm l= n(t-\tau)\,,\\
\\
J_2\equiv \mathrm G=\sqrt{ a(1-e^2)}\,,\qquad & \omega_2\equiv \mathrm g=\omega\,,\\
\\
J_1\equiv \mathrm H=\sqrt{a(1-e^2)}\,\cos i\,,\qquad & \omega_1\equiv \mathrm h=\Omega\,,
\end{array}
\]
where $n$ is the mean motion, $a$ is the semimajor axis of the orbit, $e$ is the eccentricity, $i$ is the
inclination, $\omega$ is the argument of the perigee, $\Omega$ is the longitude of the ascending
node, $\tau$τ is the time when the satellite passes through the perigee.

In the Delaunay type variables the Hamilton function $H$ (\ref{kepl-ham1}) takes the form
\bq\label{kepl-ham2}
H=-\dfrac{1}{2J_3^2}
\eq
and, therefore, we can construct the bi-Hamiltonian formulation of the perturbed Kepler model with the second bivector
$P'$ given by (\ref{part1-poi}). For instance, if
\[\beta_1(J_1)=J_1\,,\qquad \beta_2(J_2)=J_2\,,\qquad \mbox{and}\qquad K=-\dfrac{1}{3J_3^3}\,,\]
second bivector is equal to
\[
P'=\sum_{k=1}^3 J_k\,\dfrac{\partial}{\partial J_k}\wedge\dfrac{\partial}{\partial \omega_k}=
\left(
     \begin{smallmatrix}
       0 & 0 & 0 & J_1 & 0 & 0 \\
       0 & 0 & 0& 0 &  J_2&0 \\
       0 & 0 & 0 & 0 & 0 & J_3 \\
       -J_1 & 0 & 0 & 0 & 0 & 0 \\
       0 & -J_2 & 0 & 0 & 0 & 0 \\
       0 & 0 & -J_3 & 0 & 0 & 0 \\
     \end{smallmatrix}\right)\,.
\]
The corresponding recursion operator  has three functionally independent eigenvalues which are the  first integrals.

In the initial action-angle variables $(\omega_r,\omega_\theta,\omega_\phi,J_r,J_\theta,J_\phi)$ this bivector has a more complicated  form
\[
P'=\left(
     \begin{smallmatrix}
       0 & 0 & 0 & J_r+\sqrt{(J_\phi+J_\theta)^2+\epsilon\,} & 0 & 0 \\
       0 & 0 & 0& -\eta&  J_\theta+J_\phi&0 \\
       0 & 0 & 0 & -\eta& J_\theta & J_\phi \\
       -J_r-\sqrt{(J_\phi+J_\theta)^2+\epsilon\,}  &\eta & \eta & 0 & 0 & 0 \\
       0 &  -J_\theta-J_\phi & -J_\theta & 0 & 0 & 0 \\
       0 & 0 & -J_\phi & 0 & 0 & 0 \\
     \end{smallmatrix}
   \right)
\]
where
\[
\eta=\frac{\ell(\ell -J_r-\sqrt{\ell^2+\epsilon\,})}{\sqrt{\ell^2+\epsilon\,}}\,,\qquad \ell= J_\theta+J_\phi\,.
\]
In  much the same way we can obtain other bi-Hamiltonian formulations associated with the two families of the   Poincar\'{e} type action-angles variables or with other known types of the action-angle variables for the perturbed Kepler problem. Recall, for instance, that action variables at the first Poicar\'{e}  type coordinate system are equal to
\[
Z=J_\theta\,,\qquad \Gamma=J_r+\sqrt{\ell^2+\epsilon\,}-J_\phi-J_\theta\,,\qquad \Lambda=J_r+\sqrt{\ell^2+\epsilon\,}\,,
\]
see \cite{fej13} and references within.

Using similar action-angle variables for the  relativistic Kepler problem we can also obtain the non degenerate bi-Hamiltonian formulation in contradiction with the Fernandes statement in \cite{fern94}.

\section{Conclusion}
Let us duplicate textually the  Fernandes theorem from \cite{fern94}:
\vskip0.1 truecm
\par\noindent
\textbf{Theorem}: \textit{
A completely integrable Hamiltonian system is bi-Hamiltonian (satisfying
(BH)) if and only if the graph of the Hamiltonian function is a hypersurface of translation,relative to the affine structure determined by the action variables.}
\vskip0.1truecm
\par\noindent
An additional assumption (BH) is that the corresponding recursion operator $N=P'P^{-1}$ has $n$ functionally independent real eigenvalues $\lambda_1,\ldots,\lambda_n$.

This formulation of the theorem is often mentioned in modern literature, see for instance  \cite{bl,bog96,bbr,olver93,serg04, sm97a,sm97}. Nevertheless,  there is some trivial  misprint because the author omits one more necessary condition of the non degeneracy (\ref{non-deg}) of the Hamiltonian function, which could be found in the assumption "ND"\, on the page 5 in \cite{fern94} and in the proof of the theorem.

The author forgets about  this condition  simultaneously  in the formulation of the theorem and by considering  physical examples of the applicability of this theorem. Thus,  Fernandes \cite{fern94}  proclaims that perturbed Kepler problem does not have a  bi-Hamiltonian formulation in contrast with the Kepler problem, see page 13 in \cite{fern94}: \\
 \textit{
"Also we note that for the unperturbed Kepler problem ($\epsilon= 0$) the graph of the Hamiltonian is a surface of translation, and so it has a bi-Hamiltonian formulation (on the other hand, one can show that the relativistic Kepler problem also does not have a bi-Hamiltonian formulation)."}
\\
Let us repeat that initial and perturbed Kepler systems are degenerate systems and, therefore, we can not use the Fernandes theorem for both  these systems simultaneously.

\par\noindent
This work was partially supported by RFBR grant 13-01-00061 and SPbU grant 11.38.664.2013.

\end{document}